\documentclass[twoside,12pt,ppnp]{article}
\usepackage{epsfig}

\def\Journal#1#2#3#4{{#1} {#2} (#4) #3 }

\def\PRL{\em Phys. Rev. Lett.}

\def\PRC{{\em Phys. Rev.} C}

\def\ANNP{\em Ann. Phys. (N.Y.)}

\newcommand{\be}{\begin{equation}}
\newcommand{\ee}{\end{equation}}
\newcommand{\bea}{\begin{eqnarray}}
\newcommand{\eea}{\end{eqnarray}}

\begin{document}

\title{ \vspace{1cm} Dilepton production at intermediate energies 
with in-medium spectral functions of vector mesons}
\author{E.\ Santini,$^{1}$ M. D.\ Cozma,$^{1,2}$ Amand Faessler,$^1$ C.\ Fuchs,$^{1}$\\ 
M. I.\ Krivoruchenko,$^{1,3}$ B.\ Martemyanov,$^{1,3}$\\ 
\\
$^1$Institut f\"ur Theoretische Physik, Universit\"at T\"ubingen, Germany\\
$^2$National Institute for Physics and Nuclear Engineering, Magurele-Bucharest, Romania \\
$^3$Institute for Theoretical and Experimental
Physics, Moscow, Russia}
\maketitle
\begin{abstract} 
We report on a self-consistent calculation of the 
in-medium spectral functions of the $\rho$ and $\omega$ mesons at finite baryon density. 
The corresponding in-medium dilepton spectrum is generated and 
compared with HADES data. We find that an iterative calculation of the vector
meson spectral functions provides a reasonable description of the experimental data.
\end{abstract}

The in-medium properties of hadrons are generally expressed in 
terms of the self-energy which fixes the shape of the 
spectral function of the quasiparticle in the medium. To leading 
order in density, the  self-energy is determined by the forward scattering 
amplitude of the hadron with the surrounding particles. 
In Ref. \cite{Santini:2008pk}, the forward scattering of vector mesons on nucleons 
is calculated within a nucleon resonance dominance (NRD) model. The couplings of 
resonances to the nucleon and vector meson are described by 
the extended vector meson dominance (eVMD) model, extensively formulated in 
Ref. \cite{Krivoruchenko:2001jk}. In this brief contribution we report 
some of the results presented in \cite{Santini:2008pk}, to which we refer the reader 
for more exhaustive analyses and discussions.

\begin{figure}[!htb]
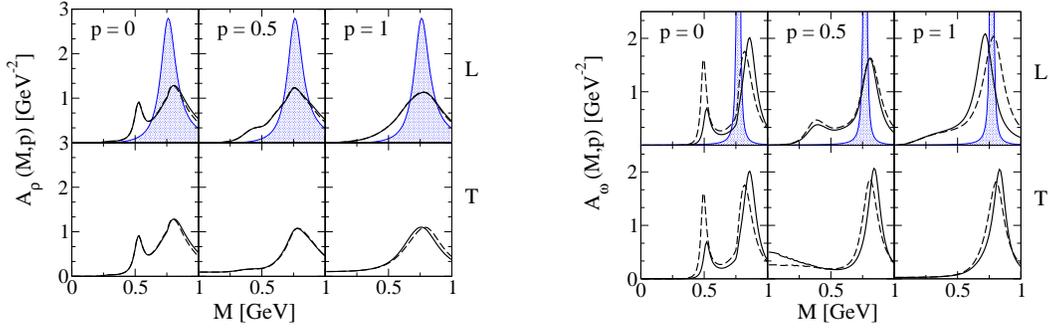

\centering
\includegraphics[width=.34\textwidth]{rhomesspecfun.eps}
\hspace{1cm}
\includegraphics[width=.34\textwidth]{omegamesspecfun.eps}
\caption{(Color online) Longitudinal ($L$) and transverse ($T$) spectral functions 
of the $\rho$ (left) and $\omega$ (right) mesons
in nuclear matter at saturation density for various momenta $p$ (in
GeV). Dashed lines stand for the resonance approximation, solid lines 
represent calculations that also included the nonresonant contributions. 
The shaded area shows the vacuum
spectral function.}
\label{rho_ome_specfun}
\end{figure}

In a first step, the in-medium vector meson self-energies are determined to leading order in density. Vacuum properties are assumed for the nucleon resonances in the calculation of the 
invariant forward scattering amplitude of vector mesons on nucleons, the latter beeing of Breit-Wigner form for resonance scattering. 
Figure~\ref{rho_ome_specfun} shows 
the resulting $\rho$ and $\omega$ spectral functions in nuclear matter at nuclear saturation density.
For both mesons we observe 
a slight upward mass shift and a substantial broadening. 
At low momenta, the spectral functions show a clear two-peak 
structure which vanishes with increasing momenta. 
The appearance of a first peak  in the spectral function 
around $0.5$-$0.55$ GeV is due to the coupling to low lying resonances. 
This first branch in the spectral distribution of the $\rho$, respectively $\omega$, meson 
is mainly generated by the $N^*(1520)$, respectively $N^*(1535)$, resonance.

\begin{figure}[!htb]
\centering
\includegraphics[width=.31\textwidth]{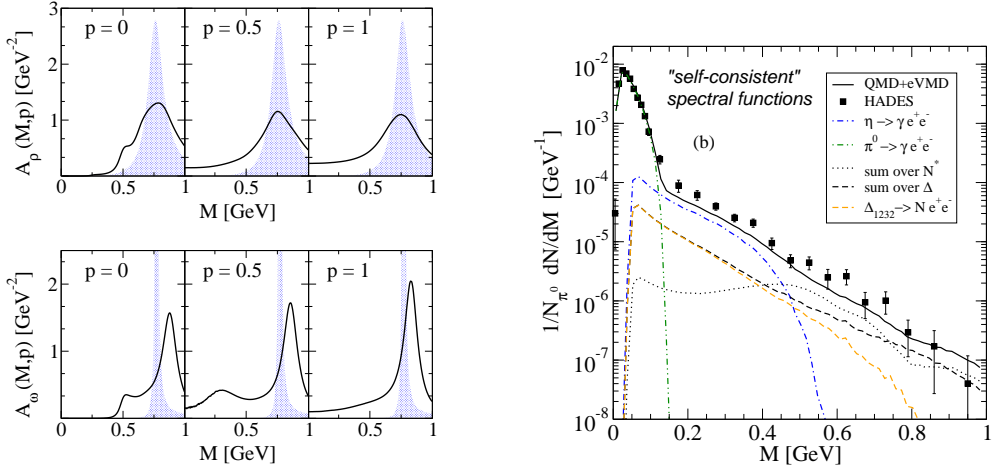}
\hspace{1cm}
\includegraphics[width=.33\textwidth]{CC2.0_s.sf_selfcon_vacres.eps}
\caption{(Color online) Dilepton spectrum in C+C collisions at 2.0$A$ GeV (right panel)
resulting from 
the inclusion of $\rho$- and $\omega$-meson spectral functions self-consistently 
calculated within the NRD+eVMD model. The latter are shown in the left panel.}
\label{rho_ome.SC}
\end{figure}

As the next step, the changes induced by the in-medium vector mesons 
on the total width of the nucleon resonances are taken into account. 
The in-medium widths of the nucleon resonances 
are determined by insertion of the in-medium spectral 
functions of the vector mesons. 
This results in a self-consistent calculation of the 
vector meson spectral functions which is solved iteratively up to convergence.
The resulting unpolarized vector meson spectral 
functions are shown in Fig.~\ref{rho_ome.SC}. 

In combination with the relativistic quantum molecular dynamics 
transport model the formalism has been applied to dilepton
emission in heavy-ion collisions at intermediate energy, 
in particular to the reaction C+C at 2$A$ GeV for which experimental data have been realised by the HADES Collaboration \cite{Agakichiev:2006tg}. 
The self-consistent 
iteration scheme provides a reasonable 
description of the data, as shown in Fig. \ref{rho_ome.SC}. 
When neglecting the in-medium properties of the nucleon resonance in the calculation of the vector meson self energies, on the contrary, no comparable agreement is found \cite{Santini:2008pk}.
This demonstrates the importance of higher order effects, i.e.,
taking in-medium modifications for the nucleon 
resonances into account when the vector meson properties are described by 
the coupling to resonance-hole states.

\clearpage

\end{document}